\documentclass[preprint,showpacs,preprintnumbers,amsmath,amssymb]{revtex4}
\usepackage{graphicx}

\newcommand\rf[1]{(\ref{eq:#1})}
\newcommand\lab[1]{\label{eq:#1}}
\newcommand\nonu{\nonumber}
\newcommand\br{\begin{eqnarray}}
\newcommand\er{\end{eqnarray}}
\newcommand\be{\begin{equation}}
\newcommand\ee{\end{equation}}

\newcommand\lb{\lbrack}
\newcommand\rb{\rbrack}

\newcommand\llb{\left\lbrack}
\newcommand\rrb{\right\rbrack}

\renewcommand\({\left(}
\renewcommand\){\right)}
\newcommand\bv{\bigm\vert}               
\newcommand\bgv{\bigg\vert}              

\newcommand\bc{\begin{center}}
\newcommand\ec{\end{center}}

\newcommand\partder[2]{\frac{{\partial {#1}}}{{\partial {#2}}}}

\renewcommand\b{\beta}

\renewcommand\d{\delta}

\newcommand\vareps{\varepsilon}
\newcommand\g{\gamma}
\newcommand\G{\Gamma}

\newcommand\h{\frac{1}{2}}
\renewcommand\k{\kappa}
\renewcommand\l{\lambda}

\newcommand\m{\mu}
\newcommand\n{\nu}

\newcommand\p{\phi}
\newcommand\vp{\varphi}
\renewcommand\P{\Phi}
\newcommand\pa{\partial}

\newcommand\pr{\prime}

\renewcommand\r{\rho}
\newcommand\s{\sigma}

\renewcommand\t{\tau}
\renewcommand\th{\theta}

\newcommand\wti{\widetilde}

\newcommand\cA{{\mathcal A}}

\newcommand\cF{{\mathcal F}}

\newcommand{\ct}[1]{\cite{#1}}
\newcommand{\bib}[1]{\bibitem{#1}}
\newcommand\PRL[3]{\textsl{Phys. Rev. Lett.} \textbf{#1}, #3 (#2)}
\newcommand\NPB[3]{\textsl{Nucl. Phys.} \textbf{B#1}, #3 (#2)}

\newcommand\PRD[3]{\textsl{Phys. Rev.} \textbf{D#1}, #3 (#2)}

\newcommand\PLB[3]{\textsl{Phys. Lett.} \textbf{#1B}, #3 (#2)}
\newcommand\CQG[3]{\textsl{Class. Quantum Grav.} \textbf{#1}, #3 (#2)}

\newcommand\PR[3]{\textsl{Phys. Reports} \textbf{#1}, #3 (#2)}

\newcommand\IJMPA[3]{\textsl{Int. J. Mod. Phys.} \textbf{A#1}, #3 (#2)}

\newcommand\MPLA[3]{\textsl{Mod. Phys. Lett.} \textbf{A#1}, #3 (#2)}

\begin{document}

\preprint{hep-th/0611022}

\title{Self-Consistent Solutions for Bulk Gravity-Matter\\
Systems Coupled to Lightlike Branes}

\author{E.I. Guendelman and A. Kaganovich}%
\email{guendel@bgumail.bgu.ac.il , alexk@bgumail.bgu.ac.il}
\affiliation{%
Department of Physics, Ben-Gurion University of the Negev \\
P.O.Box 653, IL-84105 ~Beer-Sheva, Israel
}%

\author{E. Nissimov and S. Pacheva}%
\email{nissimov@inrne.bas.bg , svetlana@inrne.bas.bg}
\affiliation{%
Institute for Nuclear Research and Nuclear Energy,
Bulgarian Academy of Sciences \\
Boul. Tsarigradsko Chausee 72, BG-1784 ~Sofia, Bulgaria
}%

\begin{abstract}
We study self-consistent $D\! =\! 4$ gravity-matter systems coupled to a new class
of Weyl-conformally invariant {\em lightlike} branes (\textsl{WILL}-branes).
The latter serve as material and charged source for gravity and electromagnetism.
Further, due to the natural coupling to a 3-index antisymmetric tensor gauge
field, the \textsl{WILL}-brane dynamically produces a {\em space-varying}
bulk cosmological constant. We find static spherically-symmetric solutions
where the space-time consists of two regions with black-hole-type geometries
separated by the \textsl{WILL}-brane which ``straddles'' their common event
horizon and, therefore, provides an explicit dynamical realization of the
``membrane paradigm'' in black hole physics. Finally, by matching via
\textsl{WILL}-brane of internal Schwarzschild-de-Sitter with external
Reissner-Nordstr{\"o}m-de-Sitter (or external Schwarzschild-de-Sitter)
geometries we discover the emergence of a potential ``well'' for infalling
test particles in the vicinity of the \textsl{WILL}-brane (the common horizon) 
with a minimum on the brane itself.

Based on talks at \textsl{IV Summer School in Modern Mathematical Physics},
Belgrade (Sept. 2006), and \textsl{2nd Workshop of European RTN} 
{\em ``Constituents, Fundamental Forces and Symmetries of the Universe''}, 
Naples (Oct. 2006).
\end{abstract}


\maketitle

\section{\label{sec:intro}Introduction}

Higher-dimensional extended objects gained in recent years a dramatically increasing
importance due to various developments in string theory, gravity, astrophysics and 
cosmology.

In non-perturbative string theory (for a background, see refs.\ct{brane-string-rev})
there arise several types of higher-dimensional 
membranes ($p$-branes, $Dp$-branes) which play a crucial role in the description of
string dualities, microscopic physics of black holes, gauge 
theory/gravity correspondence \ct{duality-AdS-CFT}, large-radius compactifications of extra 
dimensions, cosmological brane-world scenarios \ct{R-S}, 
model building in high-energy particle phenomenology \ct{brane-model-build}, 
\textit{etc.}.

Lightlike membranes are of particular interest in general relativity as they
describe impulsive lightlike signals arising in various violent astrophysical
events, \textsl{e.g.}, final explosion in cataclysmic processes such as supernovae
and collision of neutron stars \ct{barrabes-hogan}. 
Lightlike membranes are basic ingredients in the so called ``membrane paradigm'' 
theory \ct{membrane-paradigm} which appears to be a quite effective treatment of the
physics of a black hole horizon. Furthermore, the thin-wall description of domain 
walls coupled to gravity \ct{Israel-66,Barrabes-Israel-Hooft} is able to provide neat
models for many cosmological and astrophysical effects.

In refs.\ct{Israel-66,Barrabes-Israel-Hooft} lightlike membranes in the context of
gravity and cosmology have been extensively studied from a phenomenological point
of view,\textsl{i.e.}, by introducing them without specifying the Lagrangian 
dynamics from which they may originate. Recently in a series of papers
\ct{will-brane-kiten-zlatibor,will-prd} we have developed a new field-theoretic 
approach for a systematic description of the dynamics of lightlike branes starting 
from concise {\em Weyl-conformally invariant} actions. The latter are 
related to, but bear significant qualitative differences from, the standard 
Nambu-Goto-type $p$-brane actions\footnote{In ref.\ct{barrabes-israel-05} brane actions 
in terms of their pertinent extrinsic geometry have been proposed which generically
describe non-lightlike branes, whereas the lightlike branes are treated as a limiting
case.} (here $(p+1)$ is the dimension of the brane world-volume, see 
Sect.\ref{sec:will-brane} below).

Furthermore, our Weyl-conformally invariant actions are {\em not} related to the
the $p$-brane actions previously proposed in the literature \ct{Weyl-reform} where 
the standard Weyl-conformally non-invariant Nambu-Goto $p$-brane actions and their 
supersymmetric counterparts were reformulated in a formally Weyl-invariant form 
by means of introducing auxiliary
non-dynamical fields with a non-trivial transformation properties under
Weyl-conformal symmetry appropriately tuned up to compensate for the Weyl
non-invariance with respect to the original dynamical degrees of freedom.
Namely, one immediately observes that the latter formally Weyl-invariant
$p$-brane actions {\em do not change} the dynamical content of the standard
Nambu-Goto $p$-branes (describing inherently {\em massive} modes). This is in sharp
contrast to the presently discussed Weyl-conformally invariant $p$-brane models, 
which describe {\em intrinsically lightlike} $p$-branes for any even $p$
(\textsl{i.e.}, for any odd-dimensional world-volume). In what 
follows we will use for the latter the acronym \textsl{WILL-branes} 
(Weyl-invariant lightlike branes).

Our approach is based on the general idea of employing alternative non-Riemannian
integration measures (volume-forms) in the actions of generally-covariant
(reparametrization-invariant) field theories instead of (or, more generally,
on equal footing with) the standard Riemannian volume forms. Namely, instead
of (or alongside with) the standard Riemannian integration measure density
$\sqrt{-g}$ with $g = \det\Vert g_{\m\n}\Vert$ being the
determinant of the corresponding Riemannian metric, one can employ the
equally well suited non-Riemannian integration measure density:
\be
\P (\vp) \equiv \frac{1}{D!} \vareps_{i_1\ldots i_{D}}
\vareps^{m_1\ldots m_{D}} \pa_{m_1} \vp^{i_1}\ldots \pa_{m_{D}} \vp^{i_{D}}
\; ,
\lab{mod-measure-D}
\ee
where $\vp^i$ ($i=1,\ldots,D$) denote auxiliary scalar fields. This idea has
been first proposed and applied in the context of four-dimensional theories
involving gravity \ct{TMT-basic} by introducing a new class of ``two-measure''
gravitational models. It has been demonstrated that the latter models are
capable to provide plausible solutions for a broad array of basic problems
in cosmology and particle physics, such as:
(i) scale invariance and its dynamical breakdown; (ii) spontaneous generation of
dimensionful fundamental scales; (iii) the cosmological constant problem;
(iv) the problem of fermionic families; (v) applications to dark energy problem and
modern cosmological brane-world scenarios.
For a detailed discussion we refer to the series of papers \ct{TMT-basic,TMT-recent}.

Subsequently, the idea of employing an alternative non-Riemannian integration
measure was applied systematically to string, $p$-brane and $Dp$-brane models
\ct{m-string}. The main feature of these new classes of modified
string/brane theories is the appearance of the pertinent string/brane
tension as an additional dynamical degree of freedom beyond the usual string/brane
physical degrees of freedom, instead of being introduced \textsl{ad hoc} as
a dimensionful scale. The dynamical string/brane tension acquires the
physical meaning of a world-sheet electric field strength (in the string
case) or world-volume $(p+1)$-form field strength (in the $p$-brane case) and
obeys Maxwell (Yang-Mills) equations of motion or their higher-rank
anti\-symmetric tensor gauge field analogues, respectively. As a result of the
latter property the modified-measure string model with dynamical tension
yields a simple classical mechanism of ``color'' charge confinement \ct{m-string}.

The above mentioned modified-measure $p$-brane and $Dp$-brane models \ct{m-string}
share the same drawback as ordinary Nambu-Goto $p$-branes, namely that Weyl-conformal
invariance is lost beyond the simplest string case ($p\! =\! 1$).
On the other hand, the form of the action of the modified-measure
string model with dynamical tension suggested a natural way to construct explicitly
the new class of {\em Weyl-conformally invariant} $p$-brane models
{\em for any} $p$ \ct{will-brane-kiten-zlatibor,will-prd}.

The present paper has a two-fold objective. First, in Section 
\ref{sec:will-brane} we briefly review (and extend) the construction of 
\textsl{WILL-branes}, including a Kalb-Ramond-type coupling of the
latter to a space-time $(p+1)$-rank antisymmetric tensor gauge field.
In the second main part we derive systematically spherically-symmetric
solutions for the coupled system of bulk Einstein-Maxwell interacting with a 
\textsl{WILL-brane} (Sections \ref{sec:sphere-symm},\ref{sec:bulk-will-brane}), 
the latter serving as a matter and charged source and, in addition, producing 
a space-varying dynamical cosmological constant. Finally, in Section 
\ref{sec:trapping} we describe a physically interesting effect of creation of
a potential ``well'' around the \textsl{WILL-brane}, which devides space-time 
as a common horizon into two separate regions with different black hole 
geometries and whose matching is explicitly given in terms of the free 
\textsl{WILL-brane} parameters (electric charge and Kalb-Rammond coupling
constant).

\section{\label{sec:will-brane}Weyl-Conformally Invariant Lightlike Branes}

\subsection{Action and Equations of Motion. Lightlike Property}
Consider the following new kind of $p$-brane action involving {\em modified
world-volume measure} $\P (\vp)$ (cf. \rf{mod-measure-D}) and an 
{auxiliary (Abelian) world-volume gauge field $A_a$} 
\ct{will-brane-kiten-zlatibor,will-prd}:
\be
S = - \int d^{p+1}\s \,{\P (\vp)}
\Bigl\lb \h \g^{ab} \pa_a X^{\m} \pa_b X^{\n} G_{\m\n}
- {\sqrt{F_{ab}(A) F_{cd}(A) \g^{ac}\g^{bd}}}\Bigr\rb
\lab{WI-brane}
\ee
\be
{\P (\vp) \equiv \frac{1}{(p+1)!} \vareps_{i_1\ldots i_{p+1}}
\vareps^{a_1\ldots a_{p+1}} \pa_{a_1} \vp^{i_1}\ldots \pa_{a_{p+1}} \vp^{i_{p+1}}}
\lab{mod-measure-p}
\ee
Here $\g_{ab}$ denotes the intrinsic Riemannian metric on the brane
world-volume, $\g = \det\Vert\g_{ab}\Vert$,
$F_{ab} = \pa_a A_b - \pa_b A_a$ and $a,b=0,1,\ldots,p; i,j=1,\ldots,p+1$.

The above action is invariant under {Weyl (conformal) symmetry} for any $p$:
\be
\g_{ab} \longrightarrow \g^{\pr}_{ab} = \rho\,\g_{ab}  \quad ,\quad
\vp^{i} \longrightarrow \vp^{\pr\, i} = \vp^{\pr\, i} (\vp)
\lab{Weyl-conf-p}
\ee
with Jacobian 
$\det \Bigl\Vert \frac{\pa\vp^{\pr\, i}}{\pa\vp^j} \Bigr\Vert = \rho$.

Rewriting the action \rf{WI-brane} in the following equivalent form:
\be
S = - \int d^{p+1}\!\!\s \,{\chi} \sqrt{-\g}
\Bigl\lb \h \g^{ab} \pa_a X^{\m} \pa_b X^{\n} G_{\m\n}
- {\sqrt{F_{ab} F_{cd} \g^{ac}\g^{bd}}}\Bigr\rb \quad, \;\;
{\chi \equiv \frac{\P (\vp)}{\sqrt{-\g}}}
\lab{WI-brane-chi}
\ee
we see that the composite field $\chi$ plays the role of a dynamical
(variable) brane tension. Let us note the following differences of
\rf{WI-brane} (or \rf{WI-brane-chi}) w.r.t. 
the standard Nambu-Goto $p$-branes (in the Polyakov-like formulation) :
\begin{itemize}
\item
New non-Riemannian integration measure density $\P (\vp)$ instead of 
the usual $\sqrt{-\g}$, and no ``cosmological-constant'' term ($(p-1)\sqrt{-\g}$).
\item
{Variable brane tension $\chi \equiv \frac{\P (\vp)}{\sqrt{-\g}}$}
which is Weyl-conformal {\em gauge dependent}: $ \chi \to \rho^{\h(1-p)}\chi$.
\item
Auxiliary world-sheet gauge field $A_a$ in a ``square-root'' Maxwell
term \footnote{``Square-root'' Maxwell (Yang-Mills) action in
$D=4$ was originally introduced in the first ref.\ct{Spallucci}
and later generalized to ``square-root'' actions of higher-rank
antisymmetric tensor gauge fields in $D\geq 4$ in the second and
third refs.\ct{Spallucci}.}. As discussed in our previous papers on the
subject \ct{will-brane-kiten-zlatibor,will-prd}, the appearance of this
``square-root'' Maxwell term is naturally required for consistency of the
\textsl{WILL-brane} dynamics.
\item
Possibility for natural couplings of auxiliary $A_a$ to external world-volume
(``color'' charge) currents $J^a$.
\item
Weyl-invariant for {\em any} $p$; describes {{\em intrinsically
light-like} $p$-branes} for any even $p$ (\textsl{i.e.}, odd-dimensional 
world-volume). Let us also note that there are {\em NO quantum conformal 
anomalies} in {\em odd} xdimensions!.
\end{itemize}


Employing the short-hand notations:
\br
{\(\pa_a X \pa_b X\) \equiv \pa_a X^\m \pa_b X^\n G_{\m\n}\; ,\;
\sqrt{FF\g\g} \equiv \sqrt{F_{ab} F_{cd} \g^{ac}\g^{bd}}} \; ,
\lab{short-hand}
\er
the equations of motion w.r.t. measure-building auxiliary scalars $\vp^i$ 
and $\g^{ab}$ read, respectively:
\be
{\h \g^{cd}\(\pa_c X \pa_d X\) - \sqrt{FF\g\g} = M \;
\Bigl( = \mathrm{const}\Bigr)}
\lab{phi-eqs}
\ee
\be
{\h\(\pa_a X \pa_b X\) + \frac{F_{ac}\g^{cd} F_{db}}{\sqrt{FF\g\g}} = 0}
\lab{gamma-eqs}
\ee
Taking the trace in \rf{gamma-eqs} implies $M=0$ in Eq.\rf{phi-eqs}.

Next, we get the equations of motion w.r.t. auxiliary gauge field $A_a$:
\br
{\pa_b \(\frac{F_{cd}\g^{ac}\g^{bd}}{\sqrt{FF\g\g}} \P (\vp)\) = 0}
\lab{A-eqs}
\er
and the equations of motion w.r.t. $X^\m$ :
\br
{\pa_a \(\P (\vp) \g^{ab}\pa_b X^\m\) +
\P (\vp) \g^{ab}\pa_a X^\n \pa_b X^\l \G^\m_{\n\l} = 0}
\lab{X-eqs}
\er
where {$\G^\m_{\n\l}=
\h G^{\m\k}\(\pa_\n G_{\k\l}+\pa_\l G_{\k\n}-\pa_\k G_{\n\l}\)$}
is the affine connection corresponding to the external space-time
metric $G_{\m\n}$.


Consider the $\g^{ab}$-equations of motion \rf{gamma-eqs}; 
in fact, the latter are constraints analogous to the (classical) Virasoro
constraints in ordinary string theory.
Since $F_{ab}$ is anti-symmetric $(p+1)\times (p+1)$ matrix, then
{$F_{ab}$ is {\em not invertible} in any odd $(p+1)$ -- it has at
least one zero-eigenvalue vector-field $V^a$ ($F_{ab}V^b = 0$)}.
Therefore, for any odd $(p+1)$ the induced metric $\(\pa_a X \pa_b X\)$
on the world-volume of the Weyl-invariant brane is {\em singular} (as opposed to the
ordinary Nambu-Goto brane (!)) :
\br
\(\pa_a X \pa_b X\) V^b = 0 \quad ,\quad \mathrm{i.e.}\;\;
\(\pa_V X \pa_V X\) = 0 \;\; ,\;\; \(\pa_{\perp} X \pa_V X\) = 0
\lab{LL-constraints}
\er
where $\pa_V \equiv V^a \pa_a$ and $\pa_{\perp}$ are derivates along the
tangent vectors in the complement of $V^a$.

Thus, we arrive at the following important conclusion:
every point on the world-surface of the Weyl-invariant $p$-brane \rf{WI-brane}
(for odd $(p+1)$) moves with the speed of light in a time-evolution along the
zero-eigenvalue vector-field $V^a$ of $F_{ab}$. Therefore, we will name
\rf{WI-brane} (for odd $(p+1)$) by the acronym {\em WILL-brane}
(Weyl-Invariant Lightlike-brane) model.

\textbf{Remark}. In what follows we will use a natural ansatz for the 
world-volume electric field $F_{0i}=0$ implying that 
$(V^a)=(1,\underline{0})$, \textsl{i.e.}, $\pa_V = \pa_0 \equiv \pa_\t$.

\subsection{Special case $p=2$: WILL-Membrane}



Henceforth we will explicitly consider the special case $p=2$ of \rf{WI-brane},
\textsl{i.e.}, the Weyl-invariant lightlike membrane model:
\be
S = - \int d^3\s \,\P (\vp)
\Bigl\lb \h \g^{ab} \pa_a X^{\m} \pa_b X^{\n} G_{\m\n}(X)
- \sqrt{F_{ab}(A) F_{cd}(A) \g^{ac}\g^{bd}}\Bigr\rb
\lab{WILL-membrane-0}
\ee
\be
\P (\vp) \equiv \frac{1}{3!} \vareps_{ijk}
\vareps^{abc} \pa_a \vp^i \pa_b \vp^j \pa_c \vp^k \quad ,\quad 
a,b,c =0,1,2\; ,\; i,j,k=1,2,3  \; .
\lab{mod-measure-3}
\ee

Invariance under world-volume reparametrizations allows to introduce the
following standard (synchronous) gauge-fixing conditions:
\be
\g^{0i} = 0 \;\; (i=1,2) \quad ,\quad \g^{00} = -1 \; .
\lab{gauge-fix}
\ee
The residual $\t \equiv \s^0$-independent reparametrization invariance allows
for further conformally-flat gauge-fixing of the space-like part of $\g_{ab}$:
\br
\g_{ij} = a (\t,\s^1,\s^2) {\wti \g}_{ij} (\s^1,\s^2)
\lab{conf-flat}
\er
with ${\wti \g}_{ij}$ a standard reference $2D$ metric on the membrane
surface.

The ansatz $F_{0i}=0$ together with the gauge-fixed equations motion for $A_a$ \rf{A-eqs} implies:
\be
\pa_i \chi = 0  \; ,
\lab{chi-eqs}
\ee
where $\chi \equiv \frac{\P (\p)}{\sqrt{-\g}}$ (the dynamical brane tension).

Employing \rf{gauge-fix}, the remaining gauge-fixed equations of motion 
w.r.t. $\g^{ab}$ and $X^\m$ read
(recall $\(\pa_a X \pa_b X\) \equiv \pa_a X^\m \pa_b X^\n G_{\m\n}$):
\br
{\(\pa_0 X \pa_0 X\) = 0 \quad ,\quad \(\pa_0 X \pa_i X\) = 0}  \; ,
\lab{constr-0}
\er
\br
{\(\pa_i X\pa_j X\) - \h \g_{ij} \g^{kl}\(\pa_k X\pa_l X\) = 0} \; ,
\lab{constr-vir}
\er
(the latter look exactly like the classical (Virasoro) constraints for an
Euclidean string theory w.r.t. $(\s^1,\s^2)$);
\br
{\Box^{(3)} X^\m + \( - \pa_0 X^\n \pa_0 X^\l +
\g^{kl} \pa_k X^\n \pa_l X^\l \) \G^{\m}_{\n\l} = 0}  \; ,
\lab{X-eqs-3-fix}
\er
\br
{\Box^{(3)} \equiv
- \frac{1}{\chi\sqrt{\g^{(2)}}} \pa_0 \(\chi\sqrt{\g^{(2)}} \pa_0 \) +
\frac{1}{\sqrt{\g^{(2)}}}\pa_i \(\sqrt{\g^{(2)}} \g^{ij} \pa_j \)} \; .
\lab{box-3}
\er

\subsection{Coupling to Maxwell and Rank-3 Antisymmetric Tensor Gauge Field}

We can extend straightforwardly the {\em WILL}-brane model via couplings to 
{external space-time electromagnetic field $\cA_\m$} and, furthermore, to
external space-time rank 3 gauge potential $\cA_{\m\n\l}$ (Kalb-Ramond-type
coipling) keeping {\em manifest Weyl-invariance}:
\br
S = - \int d^3\s \,{\P (\vp)}
\Bigl\lb \h \g^{ab} \pa_a X^{\m} \pa_b X^{\n} G_{\m\n}
- {\sqrt{F_{ab} F_{cd} \g^{ac}\g^{bd}}}\Bigr\rb
\nonu \\
-\; q\int d^3\s \,\vareps^{abc} \cA_\m \pa_a X^\m F_{bc} \;
{
- \;\frac{\b}{3!} \int d^3\s \,\vareps^{abc} \pa_a X^\m \pa_b X^\n \pa_c X^\l \cA_{\m\n\l}}
\lab{WILL-brane+A+A3}
\er
The second Chern-Simmons-like term in \rf{WILL-brane+A+A3} is a special case 
of a class of Chern-Simmons-like couplings of extended objects to external
electromagnetic fields proposed in ref.\ct{Aaron-Eduardo}.

Let us recall the physical significance of $\cA_{\m\n\l}$ \ct{Aurilia-Townsend}. 
In $D=4$ when adding kinetic term for $\cA_{\m\n\l}$ coupled to gravity
(see Eq.\rf{E-M-WILL} below), its field-strength:
\br
\cF_{\k\l\m\n} = 4 \pa_{[\k} \cA_{\l\m\n]} = \cF \sqrt{-G} \vareps_{\k\l\m\n}
\lab{F4}
\er
with a single independent component $\cF$ produces {\em dynamical (positive)
cosmological constant}:
\be
K = 4\pi G_N \cF^2 \; .
\lab{dynamical-cc}
\ee

The constraints \rf{constr-0}--\rf{constr-vir} (gauged-fixed equations of
motion w.r.t. $\g^{ab}$ remain unaltered for the action \rf{WILL-brane+A+A3}.
Using the same gauge choice ({$\g^{0i}=0, \g^{00}=-1$}) and ansatz
for the world-volume gauge field-strength ({$F_{0i}(A) = 0$}),
the equations of motion w.r.t. $A_a$ now acquire the form:
\br
\pa_i X^\m \pa_j X^\n \cF_{\m\n}(\cA) = 0 \quad ,\quad
\pa_i \chi + \sqrt{2} q \pa_0 X^\m \pa_i X^\n \cF_{\m\n}(\cA) = 0 \; ,
\lab{A-eqs-1}
\er
(recall $\chi \equiv \frac{\P (\vp)}{\sqrt{-\g}}$ -- the brane tension,
$\cF_{\m\n} (\cA) = \pa_\m \cA_\n - \pa_\n \cA_\m$). Eqs.\rf{A-eqs-1} tell us
that consistency of charged WILL-brane dynamics implies that the
external space-time Maxwell field must have zero magnetic component normal
to the brane, as well as that the projection of the external electric field along
the brane must be proportional to the gradient of the brane tension.
Finally, the $X^\m$ equations of motion for \rf{WILL-brane+A+A3} read:
\br
{{\wti \Box}^{(3)} X^\m + \( - \pa_0 X^\n \pa_0 X^\l +
\g^{kl} \pa_k X^\n \pa_l X^\l \) \G^{\m}_{\n\l}}
\nonu \\
{-\; q \frac{\g^{kl}\(\pa_k X \pa_l X\)}{\sqrt{2}\,\chi}\,
\pa_0 X^\n \cF_{\l\n}\, G^{\l\m}
- \frac{\b}{3!} \vareps^{abc} \pa_a X^\k \pa_b X^\l \pa_c X^\n G^{\m\r}\,
\cF_{\r\k\l\n}} = 0 \; ,
\lab{X-eqs-1}
\er
where $\cF_{\r\k\l\n}$ is given as in \rf{F4} and:
\br
{{\wti \Box}^{(3)} \equiv
- \frac{1}{\chi \sqrt{\g^{(2)}}} \pa_0 \(\chi \sqrt{\g^{(2)}} \pa_0 \) +
\frac{1}{\chi \sqrt{\g^{(2)}}}\pa_i \(\chi \sqrt{\g^{(2)}} \g^{ij} \pa_j \)}
\lab{box-3-1}
\er
where $\g^{(2)} = \det\Vert\g_{ij}\Vert$ 
(recall Eqs.\rf{gauge-fix}--\rf{conf-flat}). 

As a consequence of of the constraints \rf{constr-0}--\rf{constr-vir}
and the equations of motion for $X^\m$ \rf{X-eqs-1} we also obtain:
\be
\pa_0 \(\pa_i X^\m \, \pa_j X^\n\) = 0
\lab{0-constr}
\ee

\section{\label{sec:sphere-symm}{\em WILL}-brane in Spherically-Symmetric
Backgrounds}

Let us consider the general form of spherically-symmetric gravitational 
background:
\br
{(ds)^2 = - A(r,t)(dt)^2 + B(r,t)(dr)^2 + 
C(r,t) \lb (d\th)^2 + \sin^2 (\th)\,(d\p)^2\rb}
\lab{spherical-symm-metric}
\er
where specifically:
\be
A(r) = B^{-1}(r) = 1 - \frac{2G_N M}{r}
\lab{schwarzschild}
\ee
for Schwarzschild geometry;
\be
A(r) = B^{-1}(r) = 1 - \frac{2G_NM}{r} + \frac{G_N Q^2}{r^2}
\lab{R-N}
\ee
for Reissner-Nordstr\"{o}m geometry;
\be
A(r) = B^{-1}(r) = 1 - K r^2
\lab{dS}
\ee
for (anti) de Sitter geometry;
\be
A(r) = B^{-1}(r) = 1 - K r^2 - \frac{2G_N M}{r}
\lab{schwarzschild-dS}
\ee
for Schwarzschild-(anti)-de-Sitter  geometry;
\be
A(r) = B^{-1}(r) = 1 - K r^2 - \frac{2G_N M}{r} + \frac{G_N Q^2}{r^2}
\lab{R-N-dS}
\ee
for Reissner-Nordstr\"{o}m-(anti)-de-Sitter geometry.

We will use the following ansatz:
\br
X^0 \equiv t = \t \quad ,\quad
X^1 \equiv r = r (\t,\s^1,\s^2) \quad ,\quad
X^2 \equiv \th = \s^1 \quad ,\quad X^3 \equiv \p = \s^2
\lab{so3-ansatz} \\
\g_{ij} = a (\t) \( (d\s^1)^2 + \sin^2(\s^1) (d\s^2)^2\)
\phantom{aaaaaaaaaaaaa}
\lab{gamma-ansatz}
\er
Substituting \rf{so3-ansatz}--\rf{gamma-ansatz} into the \textsl{WILL}-brane
equations of motion one gets:
\begin{itemize}
\item
Equations for $r (\t,\s^1,\s^2)$ from the lightlike \rf{constr-0}
and Virasoro-type \rf{constr-vir} constraints:
\br
{\frac{\pa r}{\pa \t} = \pm \sqrt{\frac{A}{B}}  \quad ,\quad
\frac{\pa r}{\pa \s^i} = 0}
\lab{r-eqs}
\er
\item
A strong restriction on the gravitational background itself
coming from \rf{0-constr}: 
\br
{\frac{d C}{d\t} \equiv \(\partder{C}{t} \pm
\sqrt{\frac{A}{B}}\, \partder{C}{r}\)\bgv_{t=\t,\; r=r(\t)} = 0}
\lab{C-eq}
\er
Eq.\rf{C-eq} tells us that the (squared) sphere radius $R^2 \equiv C (r,t)$ must
remain constant along the \textsl{WILL}-brane trajectory. For static backgrounds
$R^2 \equiv C(r)$ Eqs.\rf{C-eq},\rf{r-eqs} imply:
\br
{r(\t) = r_0 \;\; (= \mathrm{const})\;\; , \quad A(r_0) = 0}
\lab{will-horizon}
\er
Eq.\rf{will-horizon} is of primary importance as it shows that
{\em the \textsl{WILL}-brane automatically positions itself on the event horizon}.
\item
\textsl{WILL}-brane equations of motion \rf{X-eqs-1} for  $X^0\equiv t$ and
$X^1\equiv r$ turn out to be proportional to each other and reduce to an 
equation for the conformal factor $a(\t)$ of the internal membrane metric
\rf{gamma-ansatz} :
\br
\pa_\t (\chi\, a) + \chi\, a\;\frac{\partder{}{t}\sqrt{AB} \pm \pa_r A}{\sqrt{AB}}
\pm \chi \,\frac{\pa_r C}{\sqrt{AB}} \mp \sqrt{2}q\,\frac{C}{\sqrt{AB}}\,\cF_{0r}
\pm \b \cF C \sqrt{AB} = 0
\lab{a-eq}
\er
where $\cF$ is the independent component of the rank 4 field-strength \rf{F4}
(here again one sets at the end $t=\t,\; r=r(\t)$).
\end{itemize}

\section{\label{sec:bulk-will-brane}Bulk Gravity-Matter Coupled to {\em WILL}-brane}

\subsection{Action and Equations of Motion}

Let us now consider the following coupled Einstein-Maxwell-\textsl{WILL}-brane 
system adding also a coupling to a rank 3 gauge potential:
\br
{S = \int\!\! d^4 x\,\sqrt{-G}\,\llb \frac{R(G)}{16\pi G_N}
- \frac{1}{4} \cF_{\m\n}\cF^{\m\n} - \frac{1}{4! 2 } \cF_{\k\l\m\n}\cF^{\k\l\m\n}\rrb
+ S_{\mathrm{WILL-brane}}} \; .
\lab{E-M-WILL}
\er
Here $\cF_{\m\n} = \pa_\m \cA_\n - \pa_\n \cA_\m$,
$\cF_{\k\l\m\n} = 4 \pa_{[\k} \cA_{\l\m\n]}$ as in \rf{F4}, and the
\textsl{WILL}-brane action is the same as in \rf{WILL-brane+A+A3}:
\br
S_{\mathrm{WILL-brane}} = - \int\!\! d^3\s \,\P (\vp)
\Biggl\lb \h \g^{ab} \pa_a X^{\m} \pa_b X^{\n} G_{\m\n}
- \sqrt{F_{ab} F_{cd} \g^{ac}\g^{bd}}\,\Biggr\rb
\nonu \\
- q\int\!\! d^3\s \, \vareps^{abc} \cA_\m \pa_a X^\m F_{bc} - \frac{\b}{3!} 
\int d^3\s \,\vareps^{abc} \pa_a X^\m \pa_b X^\n \pa_c X^\l \cA_{\m\n\l}  \; .
\lab{WILL-brane+A+A3-1}
\er
The equations of motion for the \textsl{WILL}-membrane subsystem are the same as
\rf{constr-0}--\rf{constr-vir} and \rf{A-eqs-1}--\rf{X-eqs-1}, whereas the 
equations for the space-time fields read:
\br
R_{\m\n} - \h G_{\m\n} R =
8\pi G_N \( T^{(EM)}_{\m\n} + T^{(rank-3)}_{\m\n} + T^{(brane)}_{\m\n}\)
\lab{Einstein-eqs}
\er
\br
\pa_\n \(\sqrt{-G}G^{\m\k}G^{\n\l} \cF_{\k\l}\) + j^\m = 0
\lab{Maxwell-eqs}
\er
\be
\vareps^{\l\m\n\k} \pa_\k \cF + \b\, \int\! d^3\s\, \d^{(4)}(x - X(\s))
\vareps^{abc} \pa_a X^{\l} \pa_a X^{\m} \pa_a X^{\n} = 0
\lab{F4-eqs}
\ee
where in the last equation we have used relation \rf{F4}. The explicit form
of the energy-momentum tensors read:
\br
T^{(EM)}_{\m\n} = \cF_{\m\k}\cF_{\n\l} G^{\k\l} - G_{\m\n}\frac{1}{4}
\cF_{\r\k}\cF_{\s\l} G^{\r\s}G^{\k\l} \; ,
\lab{T-EM}
\er
\br
T^{(rank-3)}_{\m\n} = \frac{1}{3!}\llb \cF_{\m\k\l\r} {\cF_{\n}}^{\k\l\r} -
\frac{1}{8} G_{\m\n} \cF_{\k\l\r\s} \cF^{\k\l\r\s}\rrb = - \h \cF^2 G_{\m\n} \; ,
\lab{T-rank3}
\er
\br
T^{(brane)}_{\m\n} = - G_{\m\k}G_{\n\l}
\int\!\! d^3 \s\, \frac{\d^{(4)}\Bigl(x-X(\s)\Bigr)}{\sqrt{-G}}\,
\chi\,\sqrt{-\g} \g^{ab}\pa_a X^\k \pa_b X^\l  \; ,
\lab{T-brane}
\er
(recall $\chi \equiv \frac{\P(\vp)}{\sqrt{-\g}}$ -- the brane tension,
cf.\rf{WI-brane-chi}). The charge current in \rf{Maxwell-eqs} produced by 
the \textsl{WILL}-brane is given by:
\br
j^\m = q \int\!\! d^3 \s\,\d^{(4)}\Bigl(x-X(\s)\Bigr)
\vareps^{abc} F_{bc} \pa_a X^\m  \; .
\lab{brane-current}
\er

\subsection{Static Spherically Symmetric Solutions. Matching Accross Horizon}

We find the following static spherically symmetric solutions for the bulk
gravity-matter system coupled to a charged \textsl{WILL}-brane \rf{E-M-WILL}.
The bulk space-time consists of two regions separated by the \textsl{WILL}-brane
as a common horizon materialized by the \textsl{WILL}-brane:
\br
{(ds)^2 = - A_{(\mp)}(r)(dt)^2 + \frac{1}{A_{(\mp)}(r)}(dr)^2 +
r^2 \lb (d\th)^2 + \sin^2 (\th)\,(d\p)^2\rb}
\lab{2-regions}
\er
where the subscript $(-)$ refers to the region inside, whereas the subscript 
$(+)$ refers to the region outside the horizon at 
$r=r_0 \equiv r_{\mathrm{horizon}}$ with $A_{(\mp)}(r_0)=0$. We have
Schwarzschild-de-Sitter space-time inside horizon:
\br
A(r)\equiv A_{(-)}(r) = 1 - K_{(-)} r^2 - \frac{2G_N M_{(-)}}{r}
\;\; ,\quad \mathrm{for}\;\; r < r_0 \; ,
\lab{schwarzschild-dS-inside}
\er
and Reissner-Norstr\"{o}m-de-Sitter space-time outside horizon:
\br
{A(r)\equiv A_{(+)}(r) = 1 - K_{(+)} r^2 - \frac{2G_N M_{(+)}}{r} +
\frac{G_N Q^2}{r^2}\;\; , \quad \mathrm{for}\;\; r > r_0} \; ,
\lab{RN-dS-outside}
\er
with Reissner-Norstr\"{o}m (squared) charge given by $Q^2 = 8\pi q^2 r_{0}^4$.
The dynamically induced (due to the presence of the rank 3 tensor gauge
potential) cosmological constant is different inside and outside the horizon:
\br
K_{(\pm)} = \frac{4}{3}\pi G_N \cF_{(\pm)}^2 \quad \mathrm{for}\;\; r \geq r_{0} \;\;
(\, r \leq r_{0}\,) \quad ,\quad \cF_{(+)} = \cF_{(-)} - \b \; ,
\lab{CC-jump}
\er
where $\cF_{(\pm)}$ are the corresponding constant values of the rank 4 tensor
field-strength \rf{F4} according to Eq.\rf{F4-eqs}. 

As already discussed in Section \ref{sec:sphere-symm} 
(cf. Eqs.\rf{r-eqs}--\rf{will-horizon}), the \textsl{WILL}-membrane locates
itself automatically on (``straddles'') the common event horizon at $r=r_0$:
\br
X^0 \equiv t = \t \quad,\quad X^2 \equiv \th = \s^1 \quad,\quad 
X^3 \equiv \p = \s^2
\nonu
\er
\br
X^1 \equiv r (\t,\s^1,\s^2) = r_{0} = \mathrm{const}\;\;\mathrm{where}\;\;
A_{(-)}(r_0) = A_{(+)}(r_0) = 0
\lab{straddle}
\er

Besides inducing the jump \rf{CC-jump} in the space-varying cosmological constant,
the \textsl{WILL}-brane also causes the following discontinuity in the
normal derivative of the metric component $A(r)$ resulting from the 
\textsl{WILL}-brane contribution \rf{T-brane} to the r.h.s. of Einstein 
Eqs.\rf{Einstein-eqs}:
\br
\pa_r A_{(+)}\bv_{r=r_0} -
\pa_r A_{(-)}\bv_{r=r_0} = - 16\pi G_N \chi \; .
\lab{metric-matching}
\er
Eq.\rf{metric-matching} is easily obtained upon using the simple expressions
for the components of the Ricci tensor corresponding to the metric 
\rf{spherical-symm-metric} 
$R^0_0 = R^1_1 = -\frac{1}{2r^2} \partder{}{r}\( r^2 \partder{}{r}A(r)\)$
\ct{Eduardo-Rabinowitz} and upon substituting \rf{straddle} in \rf{T-brane}.

For the Maxwell subsystem \rf{Maxwell-eqs},\rf{brane-current} we have Coulomb 
field outside horizon generated by the surface charge of the \textsl{WILL}-brane: 
\br
{\cA_0 = \frac{\sqrt{2}\, q\, r_{0}^2}{r} \;\; ,\quad \mathrm{for}\;\;
r \geq r_{0}}
\nonu
\er
and no electric field inside horizon: 
\br
{\cA_0 = \sqrt{2}\, q\, r_{0} = \mathrm{const} \;\; ,
\quad \mathrm{for}\;\; r \leq r_{0}} \; ,
\nonu
\er

Apart from the matching conditions for the metric components \rf{metric-matching}
and the induced cosmological constant \rf{CC-jump} when crossing the
\textsl{WILL}-brane hypersurface, \textsl{WILL}-brane dynamics imposes
an important third matching condition. Namely, the remaining non-trivial
\textsl{WILL}-brane Eq.\rf{a-eq} reduces now to the following two equations
because of the two different space-time geometries inside and outside the
horizon:
\br
\pa_0 (\chi\, a) + \chi\, a \pa_r A_{(-)}\bv_{r=r_0}
- \( 2\chi r_0 + \b \cF_{(-)} r_0^2\) = 0 \; ,
\lab{a-eq-inside}
\er
\br
\pa_0 (\chi\, a) + \chi\, a \pa_r A_{(+)}\bv_{r=r_0}
- \( 2\chi r_0 + \b \cF_{(+)} r_0^2 - 2 q^2 r_0^2\) = 0 \; .
\lab{a-eq-outside}
\er
Eqs.\rf{a-eq-inside},\rf{a-eq-outside} should be yield 
the same solution for the conformal factor $a(\t)$ of the internal brane
metric, \textsl{i.e.}, we have the following additional matching condition:
\br
{\pa_r A_{(+)}\bv_{r=r_0} - \pa_r A_{(-)}\bv_{r=r_0} =
- \frac{r_0 (2q^2 + \b^2) \pa_r A_{(-)}\bv_{r=r_0} }{2\chi + \b r_0 \cF_{(-)}}}
\lab{x-eq-matching}
\er

The matching conditions \rf{metric-matching} and \rf{x-eq-matching}
allow to express all physical parameters of our solution in terms of
{3 free parameters $(q,\b,\cF)$} where:

(a) $q$ -- \textsl{WILL}-brane surface electric charge density;

(b) $\b$ -- \textsl{WILL}-brane (Kalb-Rammond-type) charge w.r.t. rank 3 
space-time gauge potential $\cA_{\l\m\n}$;

(c) $\cF_{(-)}$ -- vacuum expectation value of $\cF_{\k\l\m\n}$ \rf{F4} in
the interior region.

The corresponding explicit expressions read:
\begin{itemize}
\item
Horizon radius:
\be
r_0^2 = \frac{1}{4\pi G_N \(\cF_{(-)}^2  - \b\cF_{(-)} + q^2 +\frac{\b^2}{2}\)} \; .
\lab{r-horizon}
\ee
\item
Schwarzschild mass:
\be
M_{(-)} = \frac{r_0\,\(\frac{2}{3}\cF_{(-)}^2  - \b\cF_{(-)} + q^2 +\frac{\b^2}{2} \)}
{2G_N \(\cF_{(-)}^2  - \b\cF_{(-)} + q^2 +\frac{\b^2}{2}\)} \; .
\lab{schwarzschild-mass}
\ee
\item
Reissner-Nordstr\"{o}m mass:
\be
M_{(+)} = M_{(-)} + \frac{r_0}{2G_N \(\cF_{(-)}^2  - \b\cF_{(-)} + q^2 +\frac{\b^2}{2}\)}
\( 2q^2 + \frac{2}{3}\b\cF_{(-)} - \frac{1}{3}\b^2\) \; .
\lab{RN-mass}
\ee
\item
Reissner-Nordstr\"{o}m charge: 
\be
Q^2 = 8\pi q^2 r_0^4  \; .
\lab{RN-charge}
\ee
\item
Space-varying dynamically induced cosmological constant:
\be
K_{(\pm)} = \frac{4}{3}\pi G_N \cF_{(\pm)}^2 \quad ,\quad
\cF_{(+)} = \cF_{(-)} - \b \; .
\lab{CC-varying}
\ee
\item
Brane tension:
\be
\chi = \frac{r_0}{2}\( q^2 + \frac{\b^2}{2} - 2\b\cF_{(-)}\) \; .
\lab{brane-tension}
\ee
\end{itemize}

To determine the type of the common horizon materialized by the
\textsl{WILL}-brane, we need the expressions for the slopes of the metric
coefficients $A_{(\pm)}(r)$ at $r=r_0$. Using expressions 
\rf{r-horizon}--\rf{brane-tension} we find:
\br
\pa_r A_{(+)}\bv_{r=r_0} = - \pa_r A_{(-)}\bv_{r=r_0} \; ,
\lab{A-slopes}
\er
\br
\pa_r A_{(-)}\bv_{r=r_0} = 8\pi G_N \chi = 
4\pi G_N r_0\, \( q^2 + \frac{\b^2}{2} - 2\b\cF_{(-)}\)  \; .
\lab{A-slope-tension}
\er
Therefore, in view of Eqs.\rf{A-slopes}--\rf{A-slope-tension}
(henceforth for definiteness we assume $\b >0$ \footnote{Taking $\b<0$ does
not qualitatively change the results below. Indeed, in all relevant
expressions \rf{r-horizon}--\rf{RN-mass} and \rf{brane-tension} $\b$ appears
always multiplied by $\cF_{(-)}$. Therefore, changing the sign of $\b$ can
always be compensated by changing the sign of the 4-index field strength
which is an obvious symmetry of the action \rf{E-M-WILL}.}) :

\begin{itemize}
\item
(i) In the area of parameter space 
$\cF_{(-)} > \frac{q^2 + \frac{\b^2}{2}}{2\b}$ (\textsl{i.e.},
when $\chi < 0$ -- negative brane tension):

$~~~$
(a) the common horizon is the De-Sitter horizon from the point of view of
the interior Schwarzschild-de-Sitter geometry;

$~~~$
(b) the common horizon is the external Reissner-Nordstr\"{o}m horizon (the
larger one) from the point of view of the exterior 
Reissner-Nordstr\"{o}m-de-Sitter geometry.

The typical form of $A (r)$ is shown in Fig.1.
\item
(ii) In the opposite area of parameter space 
$\cF_{(-)} < \frac{q^2 + \frac{\b^2}{2}}{2\b}$ 
(\textsl{i.e.}, when $\chi > 0$ -- positive brane tension):

$~~~$
(a) the common horizon is the Schwarzschild horizon from the point of view of 
the internal Schwarzschild-de-Sitter geometry;

$~~~$
(b) the common horizon is internal (the smaller one) Reissner-Nordstr\"{o}m 
horizon from the point of view of the external Reissner-Nordstr\"{o}m-de-Sitter
geometry.
\end{itemize}

\begin{figure}[h]
\includegraphics{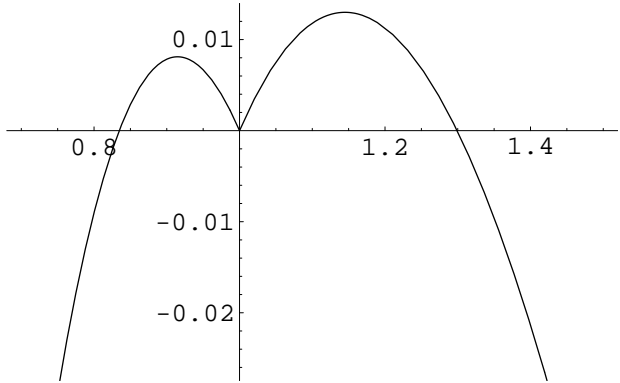}
\caption{Shape of $A (r)$ as a function of the dimensionless ratio 
$x \equiv r/r_0$}
\end{figure}

\textbf{Remark.} In the area of parameter space (i), although the brane tension
$\chi$ is negative, the brane can be stable due to negative pressure caused by 
the difference between the ``inside'' and ``outside'' dynamically induced 
cosmological constants $\cF_{(-)}>\cF_{(+)}$ (cf. Eq.\rf{CC-varying}).

In the parameter area (i), let us take the particular case $q=0$ and
$\cF_{(-)}=\b$, \textsl{i.e.}, no electromagnetic interactions present,
vanishing cosmological constant in the exterior region and the 
\textsl{WILL}-brane ``sits'' on the common horizon of an interior 
Schwarzschild-de-Sitter region 
matched with an exterior pure Schwarzschild region. The corresponding parameters 
are given as:
\be
A_{(-)}(r) = 1 - K_{(-)} r^2 - \frac{2G_N M_{(-)}}{r} \quad ,\quad
A_{(+)}(r) = 1  - \frac{2G_N M_{(+)}}{r}  \; ,
\lab{simplest-case}
\ee
for $r < r_0$ and $r > r_0$, respectively, where:
\be
r_0 = \frac{1}{\sqrt{2\pi G_N}}\, \frac{1}{\b} \quad ,\quad
K_{(-)} = \frac{4}{3}\pi G_N \b^2 \quad ,\quad
M_{(+)} = 3 M_{(-)} = \frac{1}{\sqrt{2\pi G_N}}\, \frac{1}{2\b G_N} \; .
\lab{simplest-param}
\ee
In this special case the brane tension becomes 
$\chi = - \frac{1}{\sqrt{2\pi G_N}}\, \frac{3}{4}\b$.

Let us also mention the simple special case
$\b = \cF_{(-)} = 0$,
\textsl{i.e.}, matching interior purely Schwarzschild black hole region with
exterior purely Reissner-Nordstr\"{o}m black hole region. This case has already 
been discussed in \ct{will-prd}.

The results in the present Section show that
the Einstein-Maxwell-\textsl{WILL}-brane system \rf{E-M-WILL} can be viewed as
the first explicit dynamical realization of the ``membrane paradigm'' in 
black hole physics \ct{membrane-paradigm}. Indeed, the brane dynamics
determined by the action \rf{WILL-brane+A+A3-1} dictates that the brane must
be inherently lightlike and that it necessarily has to locate itself on a
black hole event horizon. Moreover, its properties dynamically determine the
nature of the surrounding bulk gravity and matter.

\section{\label{sec:trapping}Trapping Potential Well Around Common Horizon}

Consider planar motion of a (charged) test patricle with mass $m$ and electric
charge $q_0$ in a gravitational background given by the solutions in
Section \ref{sec:bulk-will-brane}, namely, internal Schwarzschild-de-Sitter
region matched with external Reissner-Norstr{\"o}m-de-Sitter region along a 
common event horizon materialized by Weyl-conformally invariant lightlike brane,
which simultaneously serves as material and charge source for gravity and 
electromagnetism. Conservation of energy yields ($E,\, J$ -- energy and
orbital momentum of the test particle; prime indicates proper-time derivative):
\br
{\frac{E^2}{m^2} = {r^\pr}^2 + V^2_{eff}(r)}
\nonu
\er
\br
{V^2_{eff}(r) = A_{(-)}(r) \( 1 + \frac{J^2}{m^2 r^2}\)
+ \frac{2Eq_0}{m^2}\sqrt{2}q r_0 -
\frac{q_0^2}{m^2} 2q^2 r_0^2 \qquad \( r \leq r_0 \)}
\nonu \\
{V^2_{eff}(r) = A_{(+)}(r) \( 1 + \frac{J^2}{m^2 r^2}\)
+ \frac{2Eq_0}{m^2} \frac{\sqrt{2}q r_0^2}{r} -
\frac{q_0^2}{m^2} \frac{2q^2 r_0^4}{r^2} \qquad \( r \geq r_0 \)}
\er
where $A_{(\mp)}$ are the same as in \rf{schwarzschild-dS-inside} and 
\rf{RN-dS-outside}.

Taking into account \rf{A-slopes}--\rf{A-slope-tension} we see that 
in the parameter interval:
\be
\cF_{(-)} \in \(\frac{q^2 + \frac{\b^2}{2}}{\b}, \infty\)
\lab{L-interval}
\ee
the (squared) effective potential $V^2_{eff}(r)$ acquires a potential ``well''
in the vicinity of the \textsl{WILL}-brane (the common horizon) with a
minimum on the brane itself. 

\begin{figure}[b]
\includegraphics{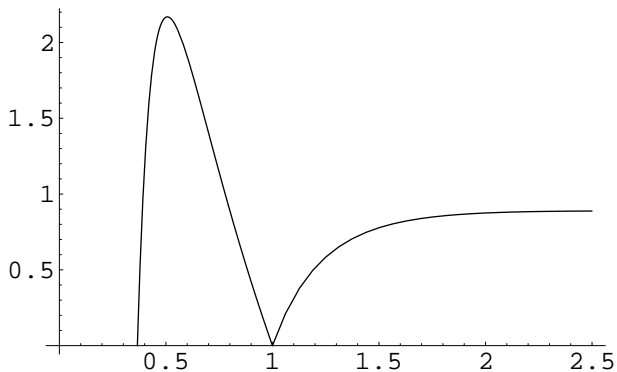}
\caption{Shape of $V^2_{eff}(r)$ as a function of the dimensionless ratio 
$x \equiv r/r_0$}
\end{figure}

Let us illustrate graphically the simplest physically interesting case
with $q=0,\, \cF_{(-)}=\b$ and $\b$ -- arbitrary,
\textsl{i.e.}, matching of Schwarzschild-de-Sitter
interior (with dynamically generated cosmological constant) 
against pure Schwarzschild exterior (with {\em no} cosmological constant)
along the \textsl{WILL}-brane as their common horizon
(cf. Eqs.\rf{simplest-case}--\rf{simplest-param}).
The typical form of $V^2_{eff}(r)$ is shown in Fig.2.

Thus, we conclude that if a test particle moving towards the common event 
horizon loses energy (\textsl{e.g.}, by radiation), it may fall and be trapped 
by the potential well, so that it neither falls into the black hole nor can
escape back to infinity. In this way one could form a ``cloud'' of trapped 
particles around the horizon.

\section{\label{sec:outlook}Conclusions and Outlook}

The present paper as well as the previous ones 
\ct{will-brane-kiten-zlatibor,will-prd} show that modifying of world-sheet 
(world-volume) integration measure (cf. Eq.\rf{mod-measure-p}) 
significantly affects string and $p$-brane dynamics.
\begin{itemize}
\item
Acceptable dynamics in the new class of string/brane models 
\rf{WI-brane} {\em naturally} requires the introduction of auxiliary
world-sheet/world-volume gauge fields.
\item
By employing square-root Yang-Mills actions for the auxiliary
world-sheet/world-volume gauge fields one achieves {{\em Weyl conformal
symmetry} in the new class of $p$-brane theories {\em for any $p$}}.
\item
Here the string/brane tension is {\em not} a constant scale given {\em ad hoc},
but rather an {\em additional dynamical degree of freedom} beyond the
ordinary string/brane degrees of freedom.
\item
Weyl-conformally invriant $p$-brane models \rf{WI-brane} describe
{\em intrinsically lightlike} $p$-branes for any even $p$
(acronym -- \textsl{WILL}-branes) with no quantum conformal anomalies
($(p+1)=$ odd).
\item
When put in a gravitational black hole background, the {\em WILL}-brane
automatically positions itself on (``materializes'') the event horizon.
\item
Coupled Einstein-Maxwell-{\em WILL}-membrane system possesses a self-consistent
solution where the {\em WILL}-membrane serves as a source for gravity and
electromagnetism. Moreover, it generates dynamical cosmological constant and
automatically ``straddles'' the common event horizon for a 
Schwarzschild-de-Sitter space-time region (in the interior) and
Reissner-Nordstr\"{o}m-de-Sitter space-time region (in the exterior).
This model is the first explicit dynamical realization of the
``membrane paradigm'' in black hole physics.
\item
It is very reasonable that negative surface tension on the horizon, produced
by the \textsl{WILL}-brane ``sitting'' on it, is the cause of a stable 
configuration where the positive cosmological constant inside the horizon is
larger than the one outside the horizon (once again due to the presence of the
\textsl{WILL}-brane). The space-varying cosmological constant causes a 
pressure difference that gives rise to an inward force on the horizon and 
this is compensated by the negative surface tension due to the 
\textsl{WILL}-brane, which pushes the horizon surface outwards.
\item
The non-trivial matching of different black hole regions along the common horizon
``straddled'' by the \textsl{WILL}-brane produces a {\em trapping potential 
``well''} for infalling test particles in the vicinity of the common horizon 
``guarding'' the inner Schwarzschild-like horizon of the interior 
Schwarzschild-de-Sitter region.
\end{itemize}

\textbf{Acknowledgements.}
{\small Two of us (E.N. and S.P.) are sincerely grateful for hospitality and
support to the organizers of the \textsl{Fourth Summer School on Modern
Mathematical Physics} (Belgrade, Serbia, Sept. 2006), and the
\textsl{Second Workshop of the European RTN} {\em ``Constituents,
Fundamental Forces and Symmetries of the Universe''} (Naples, Italy, Oct. 2006), 
where the above results were first presented. 

E.N. and S.P. are supported by European RTN networks {\em ``EUCLID''}
(contract No.\textsl{HPRN-CT-2002-00325})
and {\em ``Forces-Universe''} (contract No.\textsl{MRTN-CT-2004-005104}).
They also received partial support from Bulgarian NSF grant \textsl{F-1412/04}.
Finally, all of us acknowledge support of our collaboration through the exchange
agreement between the Ben-Gurion University of the Negev (Beer-Sheva, Israel) and
the Bulgarian Academy of Sciences.}


\end{document}